\documentclass[12pt]{article}
\newcommand{\bpi}{\mbox{\boldmath$\pi$}}
\begin{document}

\title{Area expectation values in quantum area Regge
 calculus}
\author{V.M.Khatsymovsky \\
 {\em Budker Institute of Nuclear Physics} \\ {\em
 Novosibirsk,
 630090,
 Russia}
\\ {\em E-mail address: khatsym@inp.nsk.su}}
\date{}
\maketitle
\begin{abstract}
The Regge calculus generalised to independent area tensor
variables is considered. The continuous time limit is found
and formal Feynman path integral measure corresponding to
the canonical quantisation is written out. The quantum
measure in the completely discrete theory is found which
possesses the property to lead to the Feynman path integral
in the continuous time limit whatever coordinate is chosen
as time. This measure can be well defined by passing to the
integration over imaginary field variables (area tensors).
Averaging with the help of this measure gives finite
expectation values for areas.
\end{abstract}
\newpage
The standard canonical quantisation prescription requires
continuous coordinate which would play the role of time.
Therefore this prescription is not defined in the
completely discrete theory such as Regge calculus
formulation of general relativity (GR). Due to the absence
of a unique recipe of how to quantise, a lot of different
approaches to discrete quantum gravity is possible. In
particular, a large amount of hopes are connected with the
spin foam approaches (see \cite{RegWil,Bae} for a review).
The latter are the 4D
generalisations of the 3D Ponzano-Regge model of quantum
gravity \cite{PonReg} where the partition function of 3D
Regge calculus is taken as a discrete sum of the products
of $6j$-symbols corresponding to angular momenta $j_i$ for
separate tetrahedra with linklengths $j_i+{1\over 2}$. The
basis for such a choice is that the exponential of 3D Regge
action arises in the asymptotic form of $6j$-symbols at
large $j_i$. Thus, in the spin foam models the partition
function is more fundamental object than the action since
the latter arises only in the asymptotic form of the
former. As for the standard canonical quantisation
approaches issuing directly from the action, these are
applicable to the theories with discrete space but
continuous time and are faced with the complication
consisting in the fact that discrete constraints of GR are
generally second class (i.e. do not commute). This is usual
problem with lattice regularisation in quantum gravity for
it breaks diffeomorphism invariance (see \cite{Loll} for a
review). This enforces the researchers working in this
field to develop some analogs of canonical quantisation of
general relativity for the completely discrete spacetimes
\cite{BarGamPul}.

In this letter we quantise in terms of the Feynman path
integral a model of the 4D Regge calculus \cite{Kha} which
does not
possess the above difficulty, i.e. noncommutativity of the
constraints (in the discrete space but continuous time).
Moreover, it turns out that the quantum measure for the
completely discrete theory does exist which possesses the
following property (of equivalence of the different
coordinates). If we pass to the limit when any of the
coordinates is made continuous one then we can define in
the natural way the limit for this measure which turns out
to be just the Feynman path integral measure corresponding
to the canonical quantisation with this coordinate playing
the role of time. The model in question is some
modification of the usual Regge calculus, the so-called
{\it area} Regge calculus where area tensors are treated as
independent variables. The idea that Regge calculus should
be formulated
in terms of the areas of the triangles rather than the edge
lengths was suggested by Rovelli \cite{Rov} and further
studied in ref \cite{BarRocWil}. In \cite{BarRocWil} the
area Regge calculus with independent (scalar) areas has
been introduced, and our model seems to be another one not
reducable to that of \cite{BarRocWil}. At the same time,
both versions of area Regge calculus possess simple
equations of motion which state that angle defects
vanish. Because of the lack of metric and of the usual
geometric interpretation of angle defect, this does not
mean flat spacetime. Simple form of the equations of motion
is, however, sufficient to ensure closeness of the algebra
of constraints (i.e. their being first class) even on the
discrete level. Evidently, considering the area variables
as independent ones means that the theory acquires
additional degrees of freedom. Remarkable, however, is that
this leads to simplification of the theory.

Now let us pass to the action of our specific version of
area Regge calculus of interest \cite{Kha} (in slightly
modified notations),
\begin{equation}
\label{S-VR}                                             %1
S(\pi,\Omega) = \sum_{\sigma^2}{|\pi_{\sigma^2}|\arcsin
\frac{\pi_{\sigma^2}\circ R_{\sigma^2}(\Omega)}{|\pi
_{\sigma^2}|}},
\end{equation}

\noindent where $\pi^{ab}_{\sigma^2}$ are the tensors of
the 2-faces $\sigma^2$ which in the particular case of the
usual link vector formulation (not area modified) should
reduce to $\epsilon^{abcd}l_{1c}l_{2d}$, $l^a_1$, $l^a_2$
being 4-vectors of some two edges of the triangle
$\sigma^2$ in the local frame of a certain 4-tetrahedron
containing $\sigma^2$. In our case $\pi^{ab}_{\sigma^2}$
are independent area tensors. The curvature matrix
$R_{\sigma^2}(\Omega)$ is the path-ordered product of SO(4)
(in the Euclidean case) matrices $\Omega^{\pm1}_{\sigma^3}$
living on the 3-faces $\sigma^3$ taken
along the loop enclosing the given 2-face $\sigma^2$. Some
further notations are $A\circ B$ $\equiv$ ${1\over 2}{\rm
tr}A^{\rm T}B$ for any two tensors $A$, $B$ and $|\pi
_{\sigma^2}|$ = $(\pi_{\sigma^2}\circ\pi_{\sigma^2})^{1/2}$
for twice the area of the 2-face $\sigma^2$.

The whole procedure of constructing measure includes the
following steps:\\
(i) passing to the continuous time;\\
(ii) writing out Feynman path integral following from the
Hamiltonian form and canonical quantisation;\\
(iii) finding the measure for the completely discrete
theory (including time) which would result in the above
path integral in the continuous time limit irrespectively
of what coordinate is chosen as time.

It is convenient to consider the case of the Euclidean
signature. Once the derivation is made, it turns out that
the measure obtained can be well defined by passing to the
integration over imaginary contours which looks like the
formal change of integration variables $\pi$ $\rightarrow$
$-i\pi$,
\begin{equation}                                         %2
\label{imaginary}
<f(\pi,\Omega)>=\int{f(\pi,\Omega)d\mu (\pi,\Omega)}
\Rightarrow\int{f(-i\pi,\Omega)d\mu (-i\pi,\Omega)}.
\end{equation}

\noindent It is convenient to write out integrals in such
the form from the very beginning, what just is done in the
present paper.
The considered derivation is performed quite analogously to
the case of 3D Regge calculus considered by the author in
the paper \cite{Kha1}. This is connected with the fact that
the 4D area Regge calculus resembles the 3D Regge calculus.

Passing to the continuous time limit means that some set of
the triangles, call these timelike ones, have area tensors
of the order of $O(dt)$ like
\begin{equation}
\pi_{(i^+ik)}\stackrel{\rm def}{=}n_{ik}dt.              %3
\end{equation}

\noindent Here $i^+$ means image in the leave at the moment
$t$ + $dt$ of the vertex $i$ taken at the moment $t$. (The
leaves are themselves 3D Regge manifolds.) The notation
$(i_1i_2\ldots i_{n+1})$ means unordered $n$-simplex with
vertices $i_1$, $i_2$, \ldots, $i_{n+1}$ (triangle at $n$
= 2) while that without parentheses means ordered simplex.
Besides that, the connection matrices living on the
3-tetrahedrons with volume $O(1)$ (spacelike and diagonal
ones) can be considered as those responsible for the
parallel transport at a distance $O(dt)$ and therefore it
is natural to put these being infinitesimally close to
unity. In fact, the transport to the next ($t+dt$) time
leave is defined by the sum of contributions from the
spacelike $(iklm)$ and diagonal (differing from $(iklm)$ by
occurence of superscript "+" on some of $i$, $k$, $l$, $m$)
tetrahedra inside the 4D prism with bases $(iklm)$ and
$(i^+k^+l^+m^+)$. Denote the resulting rotation matrix $1$
+ $h_{(iklm)}dt$. The considered $n_{ik}$ and $h_{(iklm)}$
turn out to be Lagrange multipliers at the constraints in
the resulting Lagrangian while area tensors on spacelike
triangles $\pi_{(ikl)}$ become dynamical variables. As for
the dynamical variables of the type of connection, these
live on the timelike tetrahedrons like $(i^+ikl)$. It turns
out, however, that dynamical connections live, in fact,
on the spacelike triangles. Indeed, consider contribution
of a diagonal triangle, e.g. $(ik^+l)$, into action. The
leading $O(1)$ contribution into action is due to the
connections on the two timelike tetrahedrons sharing this
triangle; suppose these prove to be $(ik^+kl)$ and $(ik^+
l^+l)$. Then equations of motion for the area tensor of
this triangle say that these connections are equal, may be,
up to possible inversion, $\Omega_{(ik^+kl)}$ = $\Omega
_{(ik^+l^+l)}^{\pm1}$. This means that timelike connection
is a function, in essense, on the set of bases of the 3D
prisms where the considered timelike tetrahedrons are
contained. Explicit calculation of the continuous time
action for the usual (with independent linklengths) Regge
calculus in the tetrad-connection representation has been
made by the author \cite{Kha2}. Now, in area
tensor-connection Regge calculus the problem is, in view of
the above discussion, much simpler, and expression for the
Lagrangian can be obtained from the result of \cite{Kha2}.
\begin{eqnarray}
\label{L}
L & = & L_{\dot{\Omega}}+L_h+L_n,\\                      %4
L_{\dot{\Omega}} & = & \sum_{(ikl)} \pi_{(ikl)}\circ
{\Omega}^{\dag}_{(ikl)}
\dot{\Omega}_{(ikl)},\\                                  %5
L_h & = & \sum_{(iklm)}h_{(iklm)}\circ\sum_{{\rm cycle\,
perm}\,iklm}
\varepsilon_{(ikl)m}\Omega^{\delta_{(ikl)m}}_{(ikl)}
\pi_{(ikl)}\Omega^{-\delta_{(ikl)m}}_{(ikl)}\\           %6
& \stackrel{\rm def}{=} & C(h)\nonumber\\
(\delta & \stackrel{\rm def}{=} & \frac{1+\varepsilon}{2}),
\nonumber\\
\label{L-n}
L_n & = & \sum_{(ik)}n_{(ik)}\circ R_{(ik)}\\            %7
& \stackrel{\rm def}{=} & R(n)\nonumber\\
(R_{(ik)} & = &\Omega^{\varepsilon_{ikl_n}}_{(ikl_n\!)}
\ldots\Omega^{\varepsilon_{ikl_1}}_{(ikl_1\!)},~~~
\varepsilon_{ikl_j}\!=
-\varepsilon_{(ikl_j)l_{j-1}}\!
=\varepsilon_{(ikl_j)l_{j+1}}).\nonumber
\end{eqnarray}

\noindent Here $\varepsilon_{(ikl)m}$ is a sign function
which put in correspondence +1 or -1 to each pair
of tetrahedron $(iklm)$ and triangle $(ikl)$ contained in
it. It is specified only by conditions presented in
(\ref{L-n}). The infinitesimal area tensors enter as
$n_{(ik)}$ = $n_{ik}$ + $n_{ki}$. Important property is
absence of the 'arcsin' function; this is because equations
of motion have the same solution
$R$ = $\pm 1$ as if 'arcsin' were omitted.

The eqs. (\ref{L}) - (\ref{L-n}) present the system of the
first class constraints $C$, $R$ and kinetic term analogous
to those in 3D case considered in \cite{Kha1} and first
suggested for the discrete gravity by Waelbroeck
\cite{Wae}. The difference is, first, in the local group
(SO(4) versus SO(3)) and, second, in the different topology
in 4 and 3 dimensions (here by 'topology' we mean the
scheme of connection of the different vertices); in other
respects situation is similar and, in particular,
(Euclidean) Feynman path integral measure reads
\begin{equation}
d\mu=\exp{\left(i\!\!\int{\!\!L_{\dot{\Omega}}dt}\right)}
\delta(C)\delta(R)D\pi{\cal                              %8
D}\Omega,~~~D\pi\stackrel{\rm
def}{=}\prod_{(ikl)}^{}{d^6\pi_{(ikl)}},~~~{\cal
D}\Omega\stackrel{\rm def}{=}\prod_{(ikl)}^{}{{\cal D}
\Omega_{(ikl)}}
\end{equation}

\noindent where ${\cal D}\Omega$ is the Haar measure. Upon
raising $C$, $R$ from $\delta$-functions to exponent with
the help of the Lagrange multipliers $h$, $n$ the $d\mu$
can be rewritten as
\begin{equation}
\label{d-mu}                                             %9
d\mu=\exp{\left(i\!\!\int{\!\!Ldt}\right)}D\pi Dn{\cal
D}\Omega Dh.
\end{equation}

Important is the problem of fixing the gauge and separating
out the volume of the symmetry group generated by
constraints. The gauge subgroup generated by $C$ consists
of SO(4) rotations (in the different terahedrons) and has
finite volume. Therefore there is no need to fix the
rotational symmetry. The $R(n)$ generate shifts in the
values of area tensors when $\pi_{(ikl)}$ changes by the
lateral surface of the 3D prism with the base $(ikl)$ which
is algebraic sum of $n_{(ik)}$, $n_{(kl)}$, $n_{(li)}$.
(More accurately, the sum of the expressions of the type
$\Gamma^{\dag}n\Gamma$ where $\Gamma$ are some products of
matrices $\Omega^{\pm1}$ needed to express $n$'s in the
frame of the same tetrahedron where $\pi_{(ikl)}$ is
defined.) Evidently, $R(n)$ is an analog of the Hamiltonian
constraint in the usual GR which governs the dynamics. Let
us fix the symmetry generated by $R(n)$ by fixing area
tensors of a certain set of the triangles. The number of
the triangles from this set should be the same as the
number of independent constraints $R$. The full number of
the constraints $R$ is $N^{(3)}_1$, the number of links in
the 3D leave, but $N^{(3)}_0$ of these are consequences of
others due to the Bianchi identity which can be written for
each vertex; $N^{(3)}_0$ is the number of vertices in the
3D leave. Thus, the number of independent constraints $R$
is $N^{(3)}_1$ - $N^{(3)}_0$.
Consider now {\it the set of $x$-like} triangles $F$ in the
3D leave where $x$ is some coordinate in the leave. The
number of these triangles is $N^{(3)}_1$ - $N^{(3)}_0$ as
well. More
careful investigation shows that the matrix of the Poisson
brackets $\{R,f\}$ is nondegenerate for $f$ = $\{\pi
_{(ikl)} - a_{(ikl)}|(ikl)\in F\}$, $a_{(ikl)}$ = $const$
(it's determinant is unity under
appropriate boundary conditions on the manifold) and
thereby $f$ = 0 is admissible gauge condition. By the
standard rule of the Faddeev-Popov ansatz for separating
out the gauge group volume, it is easy to find that such
gauge fixing amounts to simply omitting integrations over
$d^6\pi_{(ikl)}$, $(ikl)$ $\in$ $F$.

For subsequent constructing the completely discrete measure
respecting the coordinate equivalence it is important to
convince ourselves that analogously integration over area
tensors of the timelike triangles could be omitted. Indeed,
due to conservation in time it is sufficient to impose the
constraint $R$ only as initial condition. Therefore
integration over $Dn$ giving $\delta(R)$ in other moments
of time is not necessary and can be omitted.

As a result, one can use the measure either in the more
symmetrical form (\ref{d-mu}) or in similar form where
integration over $Dn$ is omitted or in that one where
integration over $d^6\pi_{(ikl)}$, $(ikl)$ $\in$ $F$ is
omitted, $F$ being the set of $x$-like triangles.

Now we are in a position to generalise the canonical
quantisation (continuous time) measure to the full discrete
measure. First, we can recast the measure into the
equivalent form by inserting integration over variables
living on the diagonal triangles. Let, for example, the
diagonal triangles $(ik^+l)$ and $(ik^+l^+)$ exist shared
by the tetrahedra $(ik^+kl)$, $(ik^+l^+l)$ and $(ik^+l^+
l)$, $(i^+ik^+l^+)$, respectively. Add contribution of
these triangles to the Lagrangian,
\begin{equation}
\label{diag}
\pi_{(ik^+l)}\circ\Omega^{\dag}_{(ik^+kl)}             %10
\Omega_{(ik^+l^+l)} +
\pi_{(ik^+l^+)}\circ\Omega^{\dag}_{(ik^+l^+l)}\Omega
_{(i^+ik^+l^+)},
\end{equation}

\noindent and at the same time insert integrations over
$d^6\pi_{(ik^+l)}$, $d^6\pi_{(ik^+l^+)}$ and substitute
${\cal D}\Omega_{(ikl)}$ by ${\cal D}\Omega_{(ik^+kl)}$
${\cal D}\Omega_{(ik^+l^+l)}$ ${\cal D}\Omega_{(i^+ik^+
l^+)}$. Integrations over $d^6\pi_{(ik^+l)}$, $d^6\pi_{(i
k^+l^+)}$ give $\delta$-functions of (antisymmetric parts
of) the curvature matrices which can be read off from
(\ref{diag}). These $\delta$'s are then integrated out and,
by properties of invariant Haar measure, the two additional
integrations over connections reduce to unity.

Second, it is natural to adopt the following rule of
passing from the integration over finite rotations to
that over infinitesimal ones,
\begin{equation}
{\cal D}\Omega\rightarrow d^6h,                         %11
\end{equation}

\noindent if $\Omega$ = 1 + $hdt$. Then the most
symmetrical w.r.t. the different simplices expression for
completely discrete measure should take the form
\begin{equation}
\label{d-M-F}
d{\cal M}_{\cal F} = \exp{\left (i\!\sum_{(ABC)}{\pi    %12
_{(ABC)}\circ
R_{(ABC)}(\Omega)}\right )}\prod_{(ABC)\not\in{\cal F}}d^6
\pi_{(ABC)}\prod_{(ABCD)}{{\cal D}\Omega_{(ABCD)}}.
\end{equation}

\noindent Here $A$, $B$, $C$, \ldots denote vertices of the
4D Regge manifold. Integration is omitted over area tensors
of the set of triangles ${\cal F}$. Occurrence of this set
means that full symmetry w.r.t. the different simplices is
not achieved, but a'priori there is an arbitrariness in the
choice of ${\cal F}$, so we can speak of a kind of
spontaneous symmetry breaking. Existence of this set is
connected with Bianchi identities which can lead to
singularity. Indeed, when integrating over $d^6\pi$ the
$\delta$-functions of (antisymmetric part of) the curvature
can arise, their arguments being generally not independent
just due to the Bianchi identities and possibility to have
something like $\delta$-function squared exists. Therefore
${\cal F}$ is just the set of those triangles the curvature
matrices on which are functions of other curvatures. The
number of such triangles is evidently the number of
independent Bianchi identities. Since Bianchi identity in
the 4D case can be written for the curvatures on all the
triangles sharing a given link but these identities for all
the links meeting at
a given vertex are dependent, the above number is $N^{(4)}
_1$ - $N^{(4)}_0$, $N^{(4)}_j$ being the number of
simplices of the dimensionality $j$ in the 4D manifold. The
triangles of the ${\cal F}$ should constitute a surface
which passes through all the links. Let us choose any
coordinate, denote it $t$, and consider all the $t$-like
triangles. Their number is just $N^{(4)}_1$ - $N^{(4)}_0$,
and one can prove that starting from initial conditions on
some $t$-leave one can successively express the curvatures
on these triangles in terms of other curvatures.

Further, the set ${\cal F}$ naturally fit to the
requirement to yield the canonical quantisation measure in
the continuous time limit. In fact, different choices of
${\cal F}$ correspond to different choices of gauge fixing
in the measure (\ref{d-mu}). Indeed, suppose we make the
coordinate $t$ continuous. Then, if ${\cal F}$ is the set
of the $t$-like triangles, the limit of the measure
(\ref{d-M-F}) will be (\ref{d-mu}) with integration over
area tensors $Dn$ on the timelike triangles omitted. The
choice of ${\cal F}$ being the set of $x$-like triangles
results in the limiting measure (\ref{d-mu}) with
integration over area tensors on the set $F$ $\in$ ${\cal
F}$ $x$-like triangles omitted. Finally, the choice for
${\cal F}$ being some set of the diagonal $xt$-like
triangles leads to the measure (\ref{d-mu}) with all area
tensor integrations available.

The exponential of the measure constructed containes the
terms of the two types: contributions of the $t$-like
triangles (if ${\cal F}$ is the set of $t$-like triangles
for some coordinate $t$) and contributions of the spacelike
and diagonal triangles. The former can be cast to the form
$n\circ R(R,\Omega)$ where $R(R,\Omega)$ are (rather bulky)
expressions for the curvatures on ${\cal F}$ in terms of
other curvatures and connections which solve the Bianchi
identities, but coefficients of them, tensors of the
$t$-like triangles $n$ serve as parameters and can be
chosen by hand. The latter are $\pi\circ R$ where, in
principle, $R$ can be taken as independent variables. Note
that scaling by the imaginary unity (\ref{imaginary})
refers only to the integration (dummy) variables. The
$t$-like triangle tensors $n$ enter real, and Euclidean
expectation values are defined as
\begin{eqnarray}
\label{VEV}                                             %13
<f(\pi,\Omega)> & = & \int{f(-i\pi,\Omega)\exp{\left (-\!
\sum_{\stackrel{t-{\rm like}}{(ABC)}}{n
_{(ABC)}\circ
R_{(ABC)}(\Omega)}\right )}}\nonumber\\
 & & \hspace{-20mm} \exp{\left (i
\!\sum_{\stackrel{\stackrel{\rm not}{t-{\rm like}}}{(ABC)}}
{\pi_{(ABC)}\circ
R_{(ABC)}(\Omega)}\right )}\prod_{\stackrel{\stackrel{\rm
 not}{t-{\rm like}}}{(ABC)}}d^6
\pi_{(ABC)}\prod_{(ABCD)}{{\cal D}\Omega_{(ABCD)}}.
\end{eqnarray}

\noindent Using possibility to choose tensors $n$ by hand,
take these negligibly small. If the function to be averaged
does not depend on the connections, it is easy to see, with
taking into account invariance property of the Haar
measure, that then the measure splits into the product of
the measures over separate triangles of the type
\begin{equation}
\label{separate}                                        %14
\exp{(i\pi\circ R)}d^6\pi{\cal D}R.
\end{equation}

\noindent In turn, we can use the group property SO(4) =
SO(3) $\times$ SO(3) to expand variables ($\pi$ and
generator of $R$) into self- and antiselfdual parts, in
particular, $\pi$ is mapped into 3-vectors $\,^{+}\!\bpi$,
$\,^{-}\!\bpi$.
Thereby the measure (\ref{separate}) is represented as the
product of the two measures each of which being copy of the
measure which appears in the 3D model \cite{Kha1}. In that
paper it has been found that the recipe (\ref{imaginary})
which means now 3D analog of (\ref{VEV}) indeed defines a
positive
measure if one neglects links given by hand (analogs of
$n$). Besides that, expectation value of any power $k$ $>$
-1 of the link vector squared ${\bf l}^{2k}$ turns out to
exist,
\begin{equation}                                        %15
<{\bf l}^{2k}>={4^{-k}\Gamma (2k+2)^2 \over \Gamma (k+2)
\Gamma (k+1)}
\end{equation}

\noindent (in Plank units) which extends to any function
\begin{eqnarray}
<g({\bf l})>&\!\!\!\!=&\!\!\!\!\!\int{{do_l \over       %16
4\pi}\int_{0}^{\infty}{g({\bf l})\nu (l)dl}},\nonumber\\
&&\nu (l)={2l \over \pi}\int_{0}^{\pi}{\exp{\left( -{l
\over\sin{\varphi}} \right)d\varphi}}\label{<g>}.
\end{eqnarray}

\noindent Now one should substitute here $\,^{\pm}\!\bpi$
instead of ${\bf l}$ in order to calculate expectation
values of any function of area tensor, e.g. the area itself
\begin{equation}
|\pi|^2=(\,^{+}\!\bpi)^2 + (\,^{-}\!\bpi)^2,            %17
\end{equation}

\noindent or the dual product
\begin{equation}
\label{dual}
\pi*\pi=(\,^{+}\!\bpi)^2 - (\,^{-}\!\bpi)^2.            %18
\end{equation}

\noindent The tensor $\pi$ being bivector, i.e.
antisymmetrised tensor product of two vectors, is
equivalent to the dual product vanishing. We see that this
property holds in average (but the square of (\ref{dual})
has already nonzero VEV).

The main problem is whether description in terms of lengths
can arise in the framework of the considered area
formalism. Possible approach could be to treat unambiguity
of the linklengths as a specific feature of the existing
state of the Universe. Since the question is about
unambiguity of the lengths in the full discrete spacetime,
not only in it's 3D sections, a generalisation of the usual
quantum mechanical notion of the state is implied.
Generally the measure can be viewed as a linear functional
$\mu (\Psi )$ on the (sub)space of functionals $\Psi
(\{\pi\})$ on the superspace of area tensors $\pi$ each
point of which is represented by the set of the values of
area tensors $\{\pi\}$ of all the triangles of the Regge
manifold. Of interest is some hypersurface $\Gamma$ in this
superspace which consists of the points $\{\pi\}$ defining
Regge manifolds with unambiguous lengths. Could we
consistently define our measure on the subspace of the
functionals of the form
\begin{equation}
\Psi (\{\pi\}) = \psi (\{\pi\})\delta_{\Gamma}(\{\pi\})
\end{equation}

\noindent where $\delta_{\Gamma}(\{\pi\})$ is
(many-dimensional) $\delta$-function with support on
$\Gamma$? Further, such property as positivity is required
to hold anyway to ensure probabilistic interpretation of
the measure. Therefore a strict proof is required that the
physically reasonable choice of tensors (given as
parameters) $n$ besides the trivial one $n$ = 0 considered
above exists which leaves the measure (\ref{VEV}) positive.

\bigskip

The present work was supported in part by the Russian
Foundation for Basic Research through Grant No.
01-02-16898, through Grant No. 00-15-96811 for Leading
Scientific Schools and by the Ministry of Education Grant
No. E00-3.3-148.

\end{document}